\documentclass{ws-ijmpe}
\usepackage[super,compress]{cite}
\usepackage{amssymb}
\usepackage{amsmath}
\usepackage{epstopdf}
\newcommand{\paral}{{\parallel}}
\newcommand{\vek}[1]{\bm{\mathrm{#1}}}
\newcommand{\eq}{{\mathit{eq}}}

\newcommand{\coll}{{\mathit{coll}}}

\newcommand{\WS}{{\mathit{WS}}}
\newcommand{\Ref}[1]{Ref.~\citen{#1}}
\newcommand{\Refs}[1]{Refs.~\citen{#1}}
\newcommand{\Sec}[1]{Sec.~\ref{#1}}

\newcommand{\Fig}[1]{Fig.~\ref{#1}}
\newcommand{\Eq}[1]{Eq.~(\ref{#1})}
\newcommand{\Eqs}[1]{Eqs.~(\ref{#1})}
\newcommand{\rv}{\vek{r}}

\newcommand{\qv}{\vek{q}}
\newcommand{\vv}{\vek{v}}
\newcommand{\pv}{\vek{p}}

\newcommand{\nablav}{\vek{\nabla}}

\newcommand{\xiv}{\bm{\xi}}
\newcommand{\hph}[1]{\hphantom{#1}}
\begin{document}

\markboth{M. Urban and M. Oertel}{Collective Modes in the Superfluid
  Inner Crust of Neutron Stars}

\catchline{}{}{}{}{}
%%%%%%%%%%%%%%%%%%%%%%%%%%%%%%%%%%%%%%%%%%%%%%%%%%%%%%%%%%%%%%%%%%%%

\title{Collective Modes in the Superfluid Inner Crust of Neutron Stars}

\author{Michael Urban}

\address{Institut de Physique Nucl\'eaire, CNRS-IN2P3 and Universit{\'e}
  Paris-Sud, 91406 Orsay, France\\ urban@ipno.in2p3.fr}

\author{Micaela Oertel}

\address{Laboratoire Univers et Th{\'e}ories, Observatoire de Paris,
  CNRS, and Universit{\'e} Paris Diderot, 92195 Meudon,
  France\\ micaela.oertel@obspm.fr}

\maketitle

\begin{history}
%\received{Day Month Year}
%\revised{Day Month Year}
%\accepted{Day Month Year}
%\comby{(xxxxxxxxxx)}
\end{history}

\begin{abstract}
The neutron-star inner crust is assumed to be superfluid at relevant
temperatures. The contribution of neutron quasiparticles to
thermodynamic and transport properties of the crust is therefore
strongly suppressed by the pairing gap. Nevertheless, the neutron gas
still has low-energy excitations, namely long-wavelength collective
modes. We summarize different approaches to describe the collective
modes in the crystalline phases of the inner crust and present an
improved model for the description of the collective modes in the
pasta phases within superfluid hydrodynamics.
\end{abstract}


\ccode{26.60.Gj}

%\tableofcontents

\section{Introduction}
A neutron star has a very dense core, consisting probably of
homogeneous and very neutron-rich matter (other constituents are
protons, electrons, and perhaps hyperons; the extremely dense center
of the core might even contain deconfined quark matter), which is
surrounded by the inner and the outer crust.\cite{ChamelHaensel} In
the crust, the baryon density $n_B$ is below $\sim 0.08$~fm$^{-3}$
(i.e., mass density $\varrho \lesssim 1.3\times 10^{14}$~g$/$cm$^3$)
and the matter is inhomogeneous, containing dense positively charged
``clusters'' and a highly degenerate electron gas. The difference
between the inner and the outer crust is that in the outer crust, the
clusters are simply neutron-rich nuclei, while in the inner crust (at
mass density $\varrho \gtrsim 4\times
10^{11}$~g$/$cm$^3$,\cite{ChamelHaensel} i.e., baryon density $n_B
\gtrsim 2.4\times 10^{-4}$~fm$^{-3}$), the neutron excess becomes so
strong that some neutrons are not bound any more in the clusters and
form a dilute neutron gas between the
clusters.\cite{NegeleVautherin73} To minimize Coulomb energy, the
clusters are believed to arrange in a periodic lattice. With
increasing density of the neutron gas near the core, the crystalline
lattice might transform into the so-called ``pasta'' phases, i.e.,
clusters merge first into rods (``spaghetti''), then into plates
(``lasagne''), and then some theories predict also ``inverted''
geometries where the more dilute neutron gas is concentrated in tubes
or holes (``Swiss cheese'') before the homogeneous core is
reached.\cite{Ravenhall83,Oyamatsu93}

In the first 50--100 years after the creation of the neutron star in a
supernova explosion, its core cools down very efficiently by neutrino
emission. During this so-called crust-thermalization epoch, the crust
stays hotter than the core. Since the observed temperature is that at
the surface, the thermal properties of the crust influence the
observed cooling curve.\cite{Yakovlev07} In accreting neutron stars,
the matter falling on the star results from time to time in nuclear
reactions in the crust, leading to a strong heating (released as X-ray
burst). The subsequent cooling of these X-ray transients offers
another possibility to get information on the thermodynamic properties
of the crust.\cite{Shternin2007,Brown09}

The main ingredients to describe heat transport in the crust are the
specific heat and the heat conductivity. They depend mainly on
excitations whose energy is of the same order of magnitude as the
temperature $T$. Since on a nuclear energy scale, the temperatures of
interest ($10-100$ keV, corresponding to $10^8-10^9$ K) are very low,
the most relevant excitations in the outer crust are the electrons
(specific heat $c_v \propto T$) and the phonons of the crystal lattice
($c_v\propto T^3$). In the inner crust, the situation is more
complicated, since there are in addition the unbound neutrons of the
gas between the clusters. If these neutrons were a normal Fermi gas,
their contribution to $c_v$ at low $T$ would be linear in $T$, like
that of the electrons but much larger due to the higher density of
states. However, for most of the relevant densities and temperatures
one can assume that the neutrons are superfluid (although the density
dependence of the critical temperature $T_c$ of neutron matter is not
precisely known). In this case, the energy for the creation of a
neutron quasiparticle is of the order of the pairing gap $\Delta\sim
1$~MeV, and analogously to the specific heat of a
superconductor,\cite{FetterWalecka} the contribution of neutron
quasiparticles to the specific heat is exponentially suppressed
($\propto e^{-\Delta/T}$) at low $T$. It was therefore argued that
whether neutron pairing is stronger or weaker can have an observable
effect on the cooling curve.\cite{Shternin2007,FortinMargueron} Vice
versa, observation of neutron-star cooling might help to constrain the
superfluid critical temperature in neutron star
matter.\cite{Page2010,Shternin2010}

In contrast to the gapped neutron quasiparticles, long-wavelength
collective modes of the neutron gas can be thermally excited at low
temperature and contribute to the specific heat and the heat
conductivity (although the electron and lattice-phonon contributions
are usually dominant\cite{PageReddy2012}).

This article is divided into two parts. In \Sec{sec:goldstone}, we
start by discussing collective modes in homogeneous low-density
neutron matter and in particular the appearance of the Goldstone mode
as a phase oscillation of the superfluid order parameter (gap) and its
connection with superfluid hydrodynamics. For inhomogeneous crust
matter, we then summarize results of completely microscopic
calculations, which are however limited to wavelengths smaller than
the distance between neighboring clusters, as well as results in the
long-wavelength limit. In \Sec{sec:hydropasta}, we discuss in detail a
model for collective modes in the pasta phases, in particular in the
lasagne phase, based on superfluid hydrodynamics.

Throughout the article, we use natural units with $\hbar=c=k_B=1$
($\hbar =$ reduced Planck constant, $c =$ speed of light, $k_B =$
Boltzmann constant).

%%%%%%%%%%%%%%%%%%%%%%%%%%%%%%%%%%%%%%%%%%%%%%%%%%%%%%%%%%%%%%%%%%%%%%%%%%%%%%
\section{QRPA, Goldstone Mode, and Superfluid Hydrodynamics}
\label{sec:goldstone}
%%%%%%%%%%%%%%%%%%%%%%%%%%%%%%%%%%%%%%%%%%%%%%%%%%%%%%%%%%%%%%%%%%%%%%%%%%%%%%
\subsection{QRPA in uniform neutron matter}
\label{sec:uniform}
%%%%%%%%%%%%%%%%%%%%%%%%%%%%%%%%%%%%%%%%%%%%%%%%%%%%%%%%%%%%%%%%%%%%%%%%%%%%%%
For simplicity, let us start our discussion with a uniform neutron
gas, although this is of course a very incomplete model of the inner
crust since it neglects the presence of dense clusters.

Collective modes are conveniently described in the framework of
  the Random-Phase Approximation (RPA) for the response of uniform
  matter.\cite{Garcia92,Pastore2012} It has already been used
in astrophysical contexts, e.g., to calculate the neutrino mean-free
path.\cite{Margueron2003} However, the RPA does not include pairing
and superfluidity. The extension of the RPA to systems
with pairing is called Quasiparticle RPA (QRPA).\cite{RingSchuck} QRPA
calculations in uniform neutron and neutron-star matter, using the
so-called Landau approximation for the particle-hole (ph) interaction,
were done in \Refs{Keller,BaldoDucoin2011}. Here we will discuss
results for neutron matter obtained in \Ref{MartinUrban2014} with the
full Skyrme interaction in the ph channel. Also in other fields of
physics, the RPA was generalized to systems with pairing. Let us
mention the seminal work by Anderson\cite{Anderson} and
Bogoliubov\cite{Bogoliubov} for the case of superconductors, or some
more recent work on collective modes in ultracold Fermi gases in the
unitary limit.\cite{Combescot,Forbes2013}

\begin{figure}
\parbox{6.3cm}{\includegraphics[height=6.3cm,angle=-90]{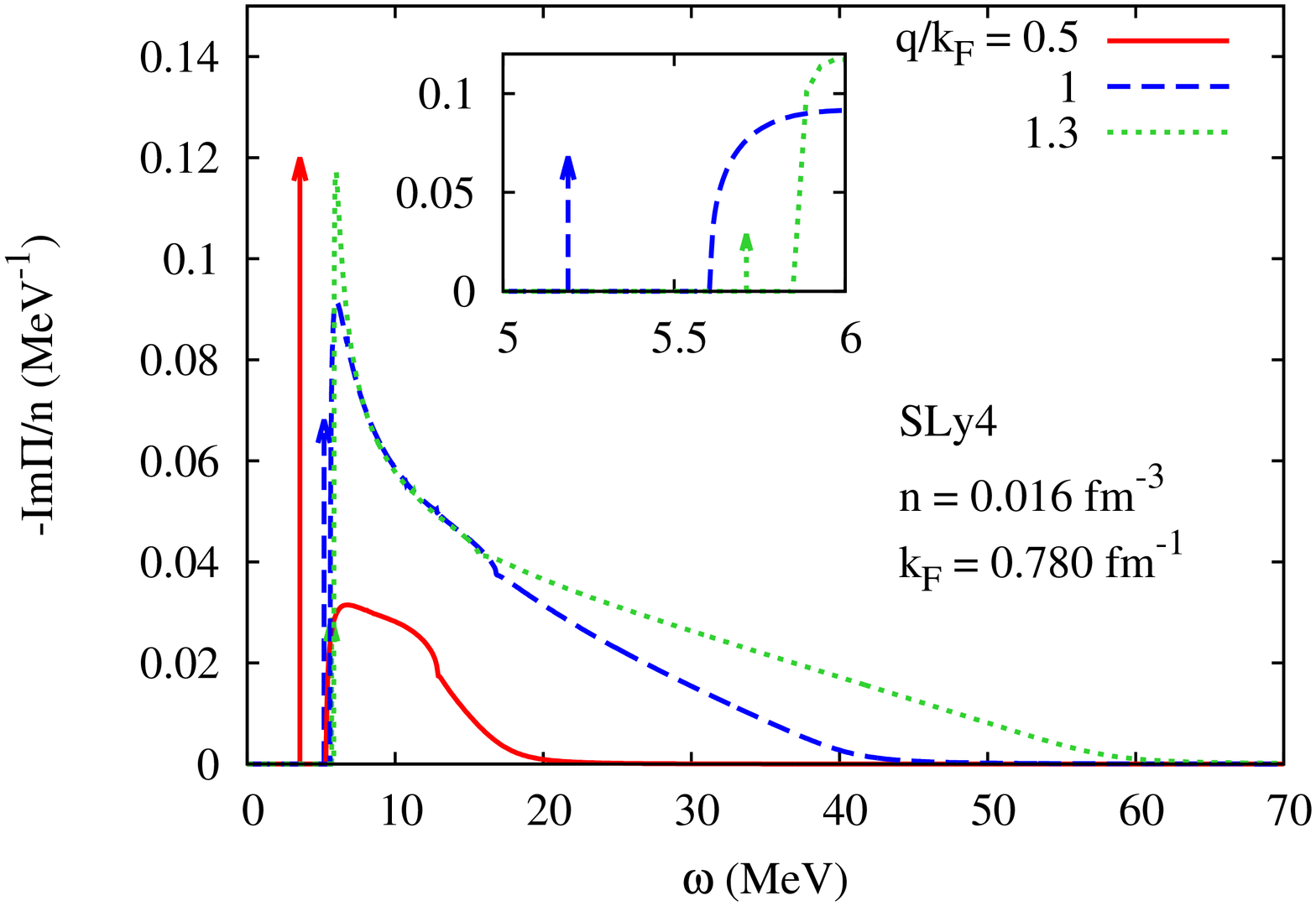}}\hfill
\parbox{6.3cm}{\includegraphics[height=6.3cm,angle=-90]{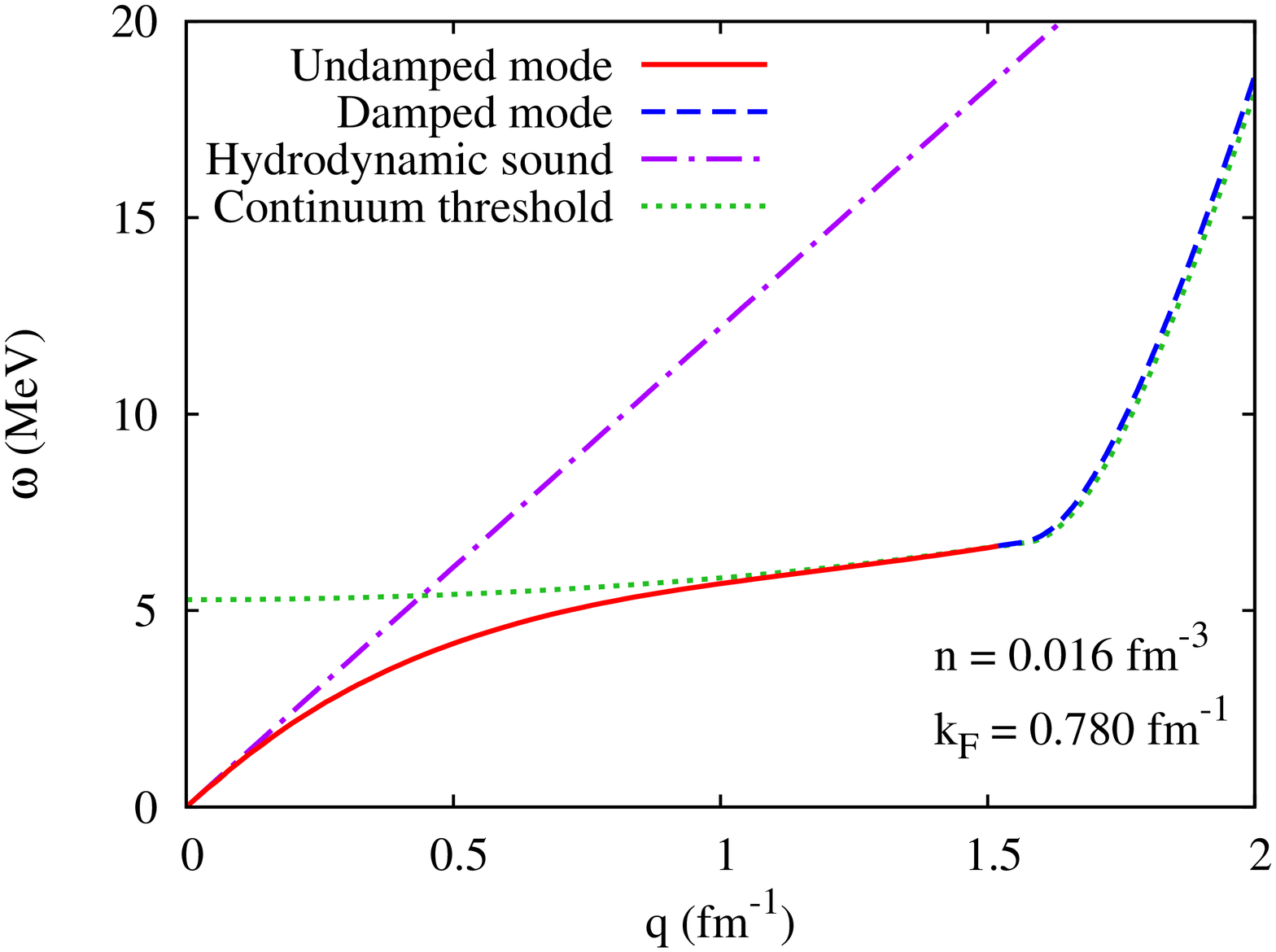}}
\caption{Left panel: QRPA results of \Ref{MartinUrban2014} for the
  density response $\Pi$ for neutron matter with density $n =
  0.016$~fm$^{-3}$. The imaginary part of $\Pi(\omega,q)$ is shown as
  function of the excitation energy $\omega$ for three different
  values of $q$. The arrows represent $\delta$-functions corresponding to
  the undamped collective mode. Right panel: Dispersion relation of
  the collective mode obtained within QRPA (solid and dashed lines)
  and the linear dispersion relation of hydrodynamic sound (dash-dotted
  line). The 2qp threshold$^{\ref{footnote:gmb}}$ is indicated by the dots.
  \label{fig:neutrongas}}
\end{figure}

The QRPA describes small oscillations around the
Hartree-Fock-Bogoliubov (HFB) ground state\footnote{In a uniform gas,
  the HFB ground state coincides with the Hartree-Fock (HF) + BCS
  one.} and can be derived by linearizing the time-dependent HFB
equations.\cite{RingSchuck} In the left panel of
\Fig{fig:neutrongas},
%%%%%%%%%%%%%%%%%%%%%%%%%%%%%%%%%%%%%%%%%%%%%%%%%%%%%%%%%%%%%%%%%%%%%%%%%%%%%%%
%%%%%%%%%%%%%%%%%%%%%%%%%%%%%%%%%%%%%%%%%%%%%%%%%%%%%%%%%%%%%%%%%%%%%%%%%%%%%%%
we show the QRPA density response function $\Pi(\omega,q)$ of neutron
matter, as a function of the excitation energy $\omega$ for different
values of the momentum $q$. At first glance, the response resembles
the broad ph continuum of the usual RPA response. Since the Landau
parameter $F_0$ relevant for the density channel is negative, there is
no zero-sound \emph{above} the continuum.\cite{Nozieres_english}
However, at low energies, one sees that, because of pairing, the ph
continuum starting at $\omega=0$ is transformed into a
two-quasiparticle (2qp) continuum starting at a threshold given by
approximately $2\Delta$.\footnote{\label{footnote:gmb}Quantitatively,
  the BCS result for $\Delta$ used in \Ref{MartinUrban2014} is too
  high since it does not account for screening of the pairing
  interaction.\cite{Gorkov1961,GezerlisCarlson2010}} This gives rise to
the appearance of an undamped collective mode \emph{below} the
threshold, indicated by the arrows.

The nature of this collective mode is a phase oscillation of the gap
$\Delta = |\Delta| e^{i\phi}$. In the ground state, the phase $\phi$
is arbitrary but constant (spontaneous $U(1)$ symmetry
breaking). Usually it is assumed without loss of generality that
$\Delta$ is real, i.e., $\phi=0$, but a global change of the phase
does not cost any energy. However, if the phase varies spatially as
$\phi \propto e^{i\qv\cdot\rv}$, the excitation energy $\omega$ is
proportional to $q$. Such a mode related to a spontaneously broken
symmetry is called a Goldstone mode,\cite{Goldstone,Weinberg2} in the
case of a superfluid it is also known as Bogoliubov-Anderson (BA)
sound. Note that this BA sound\cite{Anderson,Bogoliubov} exists only
in superfluids but not in superconductors: In the case of charged
particles, the broken symmetry is a local one (the electromagnetic
gauge symmetry) and in this case there is no Goldstone
mode.\cite{Weinberg2}

If the phase $\phi(\rv)$ is not constant, the Cooper pairs (of mass
$2m_n$, where $m_n$ denotes the neutron mass) move with a collective
velocity\cite{Landau9}
\begin{equation}
\vv = \frac{1}{2m_n}\nablav\phi\,.\label{eq:v_coll}
\end{equation}
Hence, the Goldstone mode corresponds to a longitudinal density
wave. In contrast to zero sound, the Goldstone mode does not deform
the Fermi sphere during the oscillation. Therefore its speed of sound $u$
is given by the hydrodynamic formula
\begin{equation}
  u^2 = \frac{n_n}{m_n} \frac{\partial \mu_n}{\partial n_n} \,,
\label{eq:speedofsound}
\end{equation}
where $\mu_n$ and $n_n$ are the neutron chemical potential and
density, respectively.

In the right panel of \Fig{fig:neutrongas}, we show the dispersion
relation $\omega(q)$ (solid line) of the collective mode obtained in
QRPA. As anticipated, it agrees perfectly with the hydrodynamic one
$\omega = uq$ for small $q$ (dash-dot line). However, when $\omega$
approaches the 2qp theshold (dotted line), it deviates from the linear
dispersion law. At high values of $q$, the collective mode enters the
2qp continuum and gets damped (dashed line).

At low temperatures $T\ll \Delta$, one may neglect the temperature
dependence of the dispersion relation and the damping of the Goldstone
mode. Since mainly modes with $\omega\sim T$ are excited, the
deviations from the linear dispersion law $\omega=uq$ are negligible,
too. Therefore, the neutron-gas contribution to the specific heat at
low temperature can be written analytically as
\begin{equation}
  c_{v,\coll} = \frac{2\pi^2T^3}{15\,u^3}
  \label{eq:cvPhonon}
\end{equation}
analogous to the specific heat of phonons in a
crystal.\cite{Debye,Ashcroft} On the one hand, this is of course much
larger than the contribution of neutron quasiparticles, which is
suppressed by an exponential factor $e^{-\Delta/T}$. On the other
hand, it is still much smaller than the specific heat of normal-fluid
(i.e., unpaired) neutrons, which would be linear in $T$.

%%%%%%%%%%%%%%%%%%%%%%%%%%%%%%%%%%%%%%%%%%%%%%%%%%%%%%%%%%%%%%%%%%%%%%%%%%%%%%
\subsection{QRPA in a Wigner-Seitz cell}
\label{sec:ws}
\begin{figure}
\parbox{6.1cm}{\includegraphics[width=6.1cm]{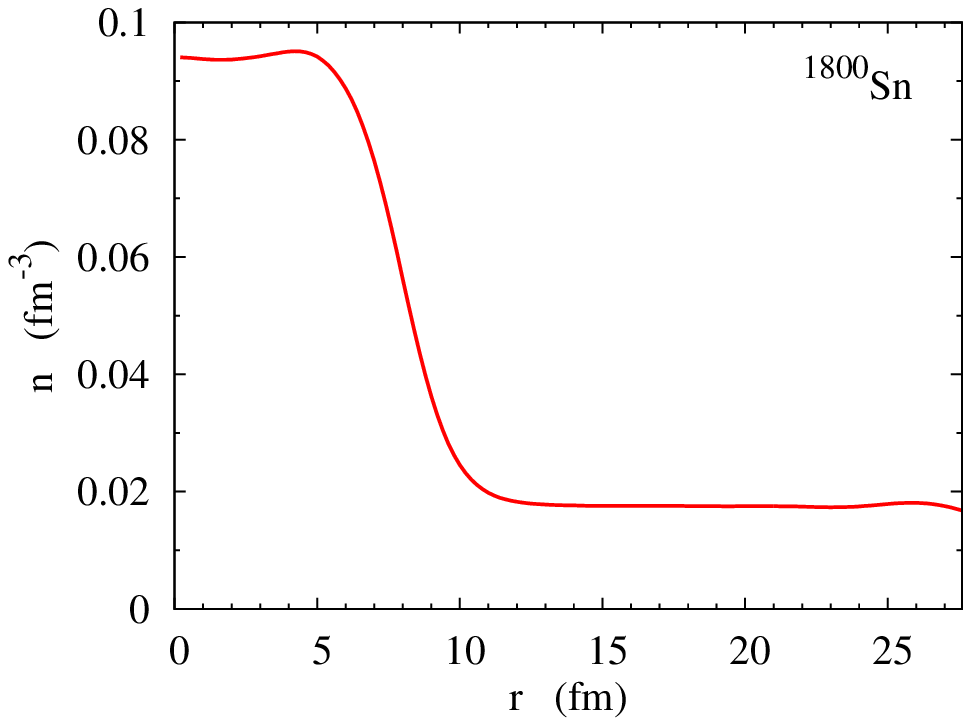}}\hfill
\parbox{6.1cm}{\includegraphics[width=6.1cm]{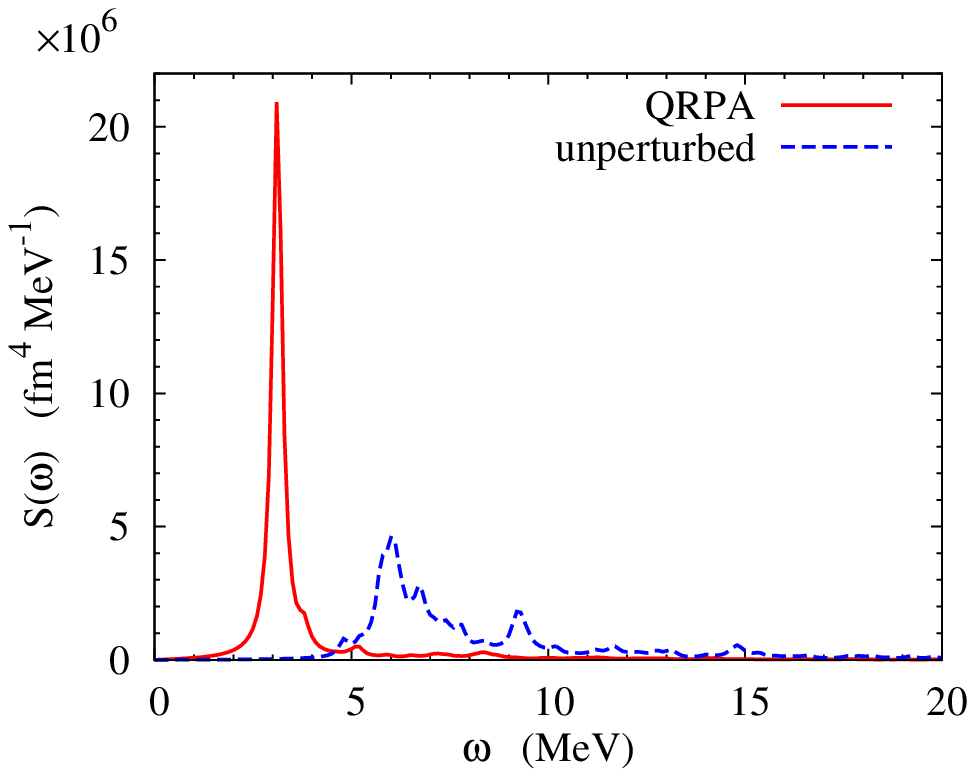}}
\caption{Left panel: HFB density profile calculated by Khan et
  al.\cite{KhanSandulescu05} in a WS cell with $R_{\WS} = 27.6$~fm,
  containing 1750 neutrons and 50 protons ($^{1800}$Sn). Right panel:
  quadrupole response of neutrons in the same WS cell calculated by
  Khan et al.\cite{KhanSandulescu05} within the QRPA (solid line; for
  comparison, the dashed line shows the unperturbed HFB quasiparticle
  response) as function of the excitation energy $\omega$.
  \label{fig:ws}}
\end{figure}

%%%%%%%%%%%%%%%%%%%%%%%%%%%%%%%%%%%%%%%%%%%%%%%%%%%%%%%%%%%%%%%%%%%%%%%%%%%%%%
So far, we have discussed only a homogeneous neutron gas. In order to
describe clusters in the gas, often the Wigner-Seitz (WS)
approximation is employed. This approximation consists in replacing
the elementary cell of the crystal lattice by a sphere of radius
$R_{\WS}$, having the same volume as the elementary cell, with the
cluster in its center. The advantage is that then all quantities,
i.e., the densities, the mean-field, the Coulomb potential, etc.,
depend only on the distance $r$ from the center of the cell. This
spherical symmetry makes it possible to carry out
HF\cite{NegeleVautherin73} or
HFB\cite{SandulescuGiai,KhanSandulescu05} calculations with the large
numbers of particles contained in a WS cell. As an example, we display
in the left panel of \Fig{fig:ws}
%%%%%%%%%%%%%%%%%%%%%%%%%%%%%%%%%%%%%%%%%%%%%%%%%%%%%%%%%%%%%%%%%%%%%%%%%%%%%%%
%%%%%%%%%%%%%%%%%%%%%%%%%%%%%%%%%%%%%%%%%%%%%%%%%%%%%%%%%%%%%%%%%%%%%%%%%%%%%%%
the HFB density profile in a WS cell containing 1750 neutrons and 50
protons ($^{1800}$Sn) calculated by Khan et al.\cite{KhanSandulescu05}
The nuclear cluster surrounded by the gas is clearly visible,
the gas density being practically constant for $r\gtrsim 12$~fm
(except for a small bump near $R_{\WS}$ which is an artefact due to
the boundary conditions\cite{NegeleVautherin73,SandulescuGiai}).

Based on the HFB ground state, Khan et al.\cite{KhanSandulescu05}
calculated the collective modes within QRPA, too. As an example, the
right panel of \Fig{fig:ws} shows the quadrupole strength (imaginary
part of the quadrupole response function) within the WS cell as a
function of the excitation energy $\omega$. A strongly collective
mode, called ``super-giant resonance'' in \Ref{KhanSandulescu05},
appears well below the unperturbed 2qp quadrupole excitations (dotted
line).

This super-giant resonance is mainly an excitation of the neutron
gas. Its energy of $\sim 3$~MeV as can be seen in \Fig{fig:ws} can be
roughly understood by considering it as the lowest eigenmode of the
hydrodynamic (BA) sound in the spherical cell. In
\Ref{KhanSandulescu05}, this energy was estimated as $\omega =
v_Fq/\sqrt{3}$ ($v_F$ being the Fermi velocity) with $q$ determined
from $j_l(qR_{\WS})=0$ ($j_l$ being a spherical Bessel function; in
the case of the quadrupole mode: $l=2$). Note, however, that
$v_F/\sqrt{3}$ is the speed of sound of an ideal Fermi gas, which is
about 50\% higher than the speed of sound $u$ one gets from
\Eq{eq:speedofsound}.

The argument for the (approximate) validity of hydrodynamics is that
the WS cell is much larger than the coherence length $\xi$ in the
neutron gas.\cite{KhanSandulescu05} Actually, up to some factors of
order unity, the condition $\xi\ll R_{\WS}$ or $\xi q\ll 1$ is equivalent
to $\omega \ll \Delta$. Obviously these conditions are not very well
satisfied, so that one expects to find at a quantitative level some
deviations from hydrodynamics, similar to the deviation of the QRPA
dispersion relation $\omega(q)$ in uniform matter from the linear one,
$\omega = uq$ (cf. \Fig{fig:neutrongas}).

Let us mention that the correspondence between QRPA and superfluid
hydrodynamics in non-uniform systems in the case of strong enough
pairing was demonstrated in \Ref{Grasso2005} in the context of
trapped ultracold Fermi gases, too. However, because of the typically
very large numbers of atoms and strong pairing in these systems, the
situation is usually much more favorable than it is in the neutron
star crust, and superfluid hydrodynamics can give precise predictions
for the frequencies of collective modes.\cite{Menotti2002}

%%%%%%%%%%%%%%%%%%%%%%%%%%%%%%%%%%%%%%%%%%%%%%%%%%%%%%%%%%%%%%%%%%%%%%%%%%%%%%
\subsection{Low-energy theory for large wavelengths}
\label{sec:eft}
%%%%%%%%%%%%%%%%%%%%%%%%%%%%%%%%%%%%%%%%%%%%%%%%%%%%%%%%%%%%%%%%%%%%%%%%%%%%%%
While the WS approximation is well suited for the description of
static properties, it is obviously not capable of describing
collective modes whose wavelengths exceed the size of the WS cell. In
reality, however, the most relevant modes at low temperature are
acoustic modes, i.e., modes whose energy $\omega$ is proportional to
$q$ for $q\ll 1/R_{\WS}$.

As mentioned in \Sec{sec:uniform}, the Goldstone mode in the uniform
neutron gas is a consequence of the broken global $U(1)$ symmetry and
corresponds to an oscillation of the phase $\phi$ of the neutron
gap. This argument remains valid in the presence of clusters. This
lead Cirigilano et al.\cite{Cirigliano2011} to develop an effective
theory for the combined system of the superfluid neutron gas and the
crystal lattice of clusters. The degrees of freedom, i.e., the fields
appearing in the effective Lagrangian, are the phase $\phi(\rv,t)$ of
the neutron gap and the displacement $\xiv(\rv,t)$ of the
clusters. Similar low-energy theories were subsequently used in
\Refs{ChamelPage2013,Kobyakov2013}. Since $\phi$ and $\xiv$ are
coarse-grained over regions larger than the periodicity $L$ of the
crystal lattice, these low-energy theories are only valid for $q\ll
1/L$.

Of course, clusters and neutron superfluid do not move independently
of each other. This is called the entrainment
effect.\cite{CarterChamel2005} It should be noted that in contrast to
an ordinary dragging of the gas by the clusters (and vice versa), the
entrainment is a non-dissipative force. It modifies the velocities
$u_l$ and $u_t$ of the longitudinal and transverse lattice phonons,
respectively, as well as the velocity $u_\phi$ of the Goldstone mode
(also called superfluid phonon).\cite{ChamelPage2013} Furthermore, the
longitudinal lattice phonons and the Goldstone mode get
mixed\cite{Cirigliano2011,ChamelPage2013,Kobyakov2013} and one finds
two new eigenmodes with two new velocities $u_{\pm}$. The framework of
low-energy effective theory was not only used to describe the
long-wavelength phonons, but also phonon-phonon and phonon-electron
interactions. These are particularly relevant for the calculation of
the phonon (lattice and superfluid) contribution to transport
properties such as the heat conductivity.\cite{PageReddy2012}

\begin{figure}
\parbox{6.1cm}{\includegraphics[width=6.1cm]{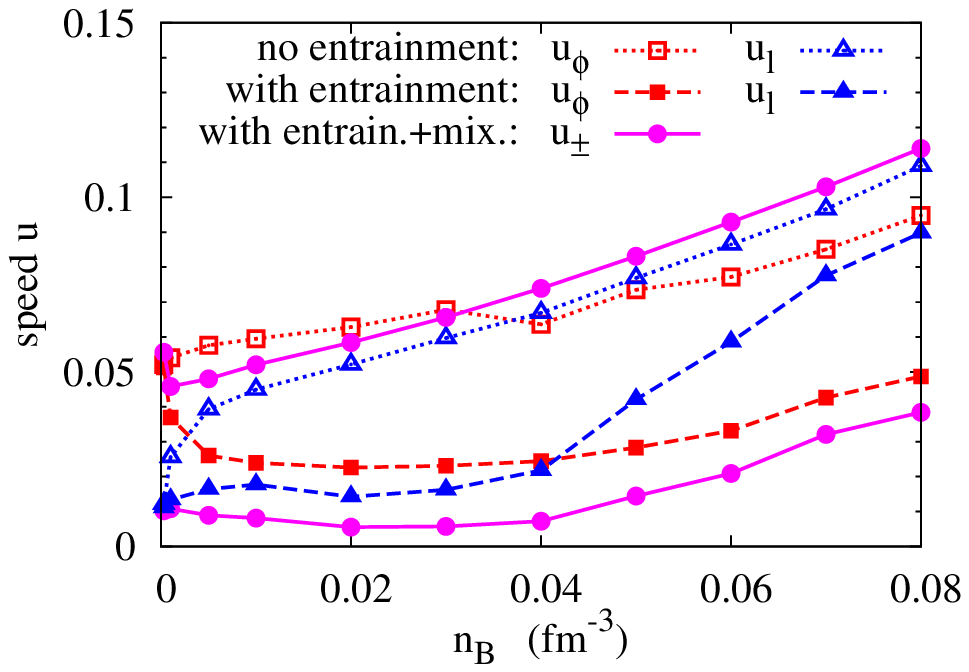}}\hfill
\parbox{6.1cm}{\includegraphics[width=6.1cm]{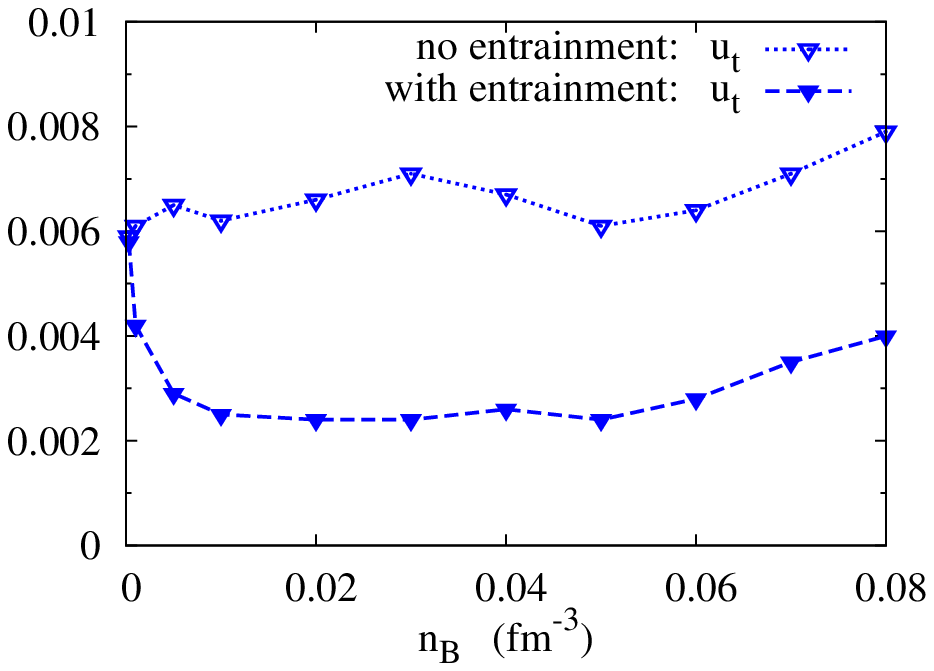}}
\caption{Velocities of the longitudinal and superfluid phonons ($u_l$
  and $u_\phi$, left panel) and of the transverse phonons ($u_t$,
  right panel) as a function of the baryon density $n_B$, from Chamel
  et al.\cite{ChamelPage2013} Dotted lines: entrainment neglected;
  dashed lines: entrainment included; solid lines: entrainment and
  mixing included.
  \label{fig:phononmix}}
\end{figure}

The coefficients of the effective theory, in particular the
entrainment and mixing coefficients, were calculated by
Chamel\cite{Chamel2012} using a band-structure calculation for the
neutrons in the periodic mean field of the clusters\cite{Chamel2005}
analogous to those used in solid-state physics for the electrons in
the periodic Coulomb field of the ions in a metal. As an example, we
display in \Fig{fig:phononmix}
%%%%%%%%%%%%%%%%%%%%%%%%%%%%%%%%%%%%%%%%%%%%%%%%%%%%%%%%%%%%%%%%%%%%%%%%%%%%%%%
%%%%%%%%%%%%%%%%%%%%%%%%%%%%%%%%%%%%%%%%%%%%%%%%%%%%%%%%%%%%%%%%%%%%%%%%%%%%%%%
the results of \Ref{ChamelPage2013} for the density dependence of the
different velocities without (dotted lines) and with (dashed lines)
entrainment and with both entrainment and mixing (solid lines). One
observes that the coupling between the lattice and the superfluid
neutron gas has a drastic effect on the lattice phonons. The
entrainment results in a reduction of the phonon velocities, since
some neutrons of the gas move together with the clusters,
\emph{increasing} the cluster effective mass (in the language of
\Refs{Chamel2012,ChamelPage2013}, the density of ``conduction
neutrons'' moving independently of the clusters is smaller than the
density of ``unbound'' neutrons).

However, as pointed out by Kobyakov and Pethick,\cite{Kobyakov2013}
zero-point oscillations of the clusters may reduce the band structure
effects. Furthermore, we note that, using a completely different
approach (similar to the one we will use in the next section),
Magierski and Bulgac\cite{Magierski2004,MagierskiBulgac2004} arrived
at the opposite result, namely that the effective mass of the clusters
is \emph{reduced} when they are immersed in the neutron gas. We
conclude that the microscopic modeling of the coefficients of the
effective theory, in particular those depending on the entrainment, is
not yet completely settled.

%%%%%%%%%%%%%%%%%%%%%%%%%%%%%%%%%%%%%%%%%%%%%%%%%%%%%%%%%%%%%%%%%%%%%%%%%%%%%%
\section{Hydrodynamic Model for Collective Modes in the Pasta Phases}
\label{sec:hydropasta}
%%%%%%%%%%%%%%%%%%%%%%%%%%%%%%%%%%%%%%%%%%%%%%%%%%%%%%%%%%%%%%%%%%%%%%%%%%%%%%
After this survey of works pointing out the connection between the
Goldstone mode of the superfluid neutrons and superfluid
hydrodynamics, let us now describe in some detail our hydroynamic
model for the collective modes, extending our preceding
work.\cite{DiGallo2011} In particular, we include the Coulomb
interaction, neglected in \Ref{DiGallo2011}, in order to describe
simultaneously superfluid modes and lattice phonons. Our aim is to
cover also wavelengths that are large compared to the coherence length
(cf. discussion in \Sec{sec:ws}) but not necessarily large compared to
the periodicity $L$ of the crystal lattice. In this sense the approach
bridges between the two extreme cases discussed above. In principle
the formalism can be applied to all phases of the inner crust, but in
the practical calculations we restrict ourselves to the simplest case,
which is the phase of plates (``lasagne''), see
Sec.~\ref{sec:modesinlasagne}.
%%%%%%%%%%%%%%%%%%%%%%%%%%%%%%%%%%%%%%%%%%%%%%%%%%%%%%%%%%%%%%%%%%%%%%%%%%%%%%
\subsection{Nuclear pasta as phase coexistence in equilibrium}
\label{sec:equilibrium}
%%%%%%%%%%%%%%%%%%%%%%%%%%%%%%%%%%%%%%%%%%%%%%%%%%%%%%%%%%%%%%%%%%%%%%%%%%%%%%
For the ground-state configuration, we use a very simple model. We
describe matter in the inner crust as a mixed phase consisting of a
neutron gas (phase 1) and a neutron-proton liquid (phase 2). We assume
that in equilibrium the densities in each phase are constant and the
two phases are separated by a sharp interface. We neglect the smooth
variation of the density, found in microscopic calculations (cf. left
panel of \Fig{fig:ws}), at the transition from one phase to the
other. Since the phases coexist, the pressure and the chemical
potentials must be equal in both phases. In addition, an electron gas
globally compensates the charge of the protons. We approximate it by
an ideal gas of massless fermions with uniform density. Since
neutrons, protons and electrons are in $\beta$-equilibrium, their
chemical potentials satisfy $\mu_{n1} = \mu_{n2} = \mu_{p2}+\mu_e$\,.
Obviously the model as stated above is very crude. In order to account
at least in some approximate way for the realistic (smooth) interface,
we assume that its effect can be subsumed in a single parameter, the
surface tension $\sigma$ (energy per interface area).

%%%%%%%%%%%%%%%%%%%%%%%%%%%%%%%%%%%%%%%%%%%%%%%%%%%%%%%%%%%%%%%%%%%%%%%%
\begin{figure}
\centerline{\includegraphics[width=4cm]{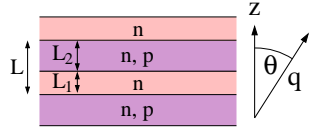}}
\caption{Periodic slab structure as simplified model for the lasagne phase.
  \label{fig:periodic}}
\end{figure}
%%%%%%%%%%%%%%%%%%%%%%%%%%%%%%%%%%%%%%%%%%%%%%%%%%%%%%%%%%%%%%%%%%%%%%%%
%%%%%%%%%%%%%%%%%%%%%%%%%%%%%%%%%%%%%%%%%%%%%%%%%%%%%%%%%%%%%%%%%%%%%%%%%%%%%%
\subsection{Hydrodynamic equations with Coulomb potential}
\label{sec:hydro}
%%%%%%%%%%%%%%%%%%%%%%%%%%%%%%%%%%%%%%%%%%%%%%%%%%%%%%%%%%%%%%%%%%%%%%%%%%%%%%
Since neutrons and protons are paired, we assume that their motion can
be described by superfluid hydrodynamic equations, except at the
interface between the two phases, where appropriate boundary
conditions are needed. This is similar to the approach used in
\Refs{Magierski2004,MagierskiBulgac2004} to calculate the flow of the
neutron gas around and through a moving cluster and the corresponding
effective mass of the cluster.

In superfluid hydrodynamics, the momentum per particle of each fluid,
$\pv_a$ ($a = n,p$ indicating neutrons or protons), is related to the
phase $\phi_a$ of the corresponding superfluid order parameter by
\begin{equation}
\label{eq:velocitypotential}
\pv_a(\rv,t) = \frac{1}{2}\nablav\phi_a(\rv,t)\,.
\end{equation}

In the neutron gas, the momentum $\pv_n$ is proportional to the
velocity $\vv_n$ and we may write
\begin{equation}
\pv_n = M_n \vv_n\,.
\label{eq:entrainment_n}
\end{equation}
In the non-relativistic limit, i.e., if the total energy density is
dominated by the mass density, the proportionality constant $M_n$ is
of course given by the neutron mass $m_n$, and one recovers the
well-known expression (\ref{eq:v_coll}) for the superfluid
velocity. Relativistically, however, one finds\cite{Landau6} that
$M_n$ is equal to the enthalpy per particle, which, at zero
temperature, equals $\mu_n$.

In the liquid phase containing neutrons and protons, the situation is
more complicated. If both fluids have different velocities
$\vv_n\neq\vv_p$, the interaction between neutrons and protons can
give rise to entrainment, i.e., to a misalignment between momenta and
velocities,\cite{Andreev1975} even in a uniform system and already at
a microscopic level (i.e., independently of the macroscopic
entrainment effect mentioned in \Sec{sec:eft}, which is a consequence
of the crystalline structure). This microscopic entrainment effect can
easily be understood in the framework of Landau Fermi-liquid
theory.\cite{Borumand1996,Chamel2006} The relationship between
velocities and momenta can then be written as
\begin{equation}
\pv_a = \sum_{b=n,p} M_{ab} \vv_b\,,
\label{eq:entrainment_np}
\end{equation}
where the matrix $M$ is related to the so-called Andreev-Bashkin
entrainment matrix, $Y$ in the notation of \Ref{Gusakov2009}, via
$M_{ab} = \sum_b Y^{-1}_{ab} n_b$. In the non-relativistic case,
Galilean invariance implies that $\sum_b M_{ab} = m_a$, and in the
absence of entrainment, the matrix $M$ reduces to $M_{ab} = m_a
\delta_{ab}$. In the relativistic case, $M$ satisfies $\sum_b M_{ab} =
\mu_a$ (see \ref{sec:microscopicinput}).

After these preliminary remarks, let us turn to the hydrodynamic
description of small oscillations. The first hydrodynamic equation is
the continuity equation, describing the conservation of particle
number of each species $a$. Since we are only interested in small
oscillations, we consider the fluid velocities $\vv_a$ as small. In
addition, we decompose the density into its constant equilibrium value
$n_{a,\eq}$ and a small deviation $\delta n_a(\rv,t) =
n_a(\rv,t)-n_{a,\eq}$. Keeping only terms linear in the small
deviations from equilibrium, we can write the continuity equation for
species $a$ as
\begin{equation}
\label{eq:continuity}
\partial_t \delta n_a(\rv,t) = -n_{a,\eq}\nablav\cdot\vv_a(\rv,t)\,.
\end{equation}
For the sake of better readability, we will from now on drop
the index $\eq$ since it is clear that, after linearization, any
quantity that is multiplied by a small quantity (such as $\vv_a$ or
$\delta n_a$) has to be replaced by its equilibrium value.

The second equation is the Euler equation, describing the conservation
of momentum. Keeping again only terms linear in the deviations from
equilibrium, we obtain in the case of the neutron gas
\begin{equation}
\partial_t \pv_n(\rv,t) =
-\frac{\partial\mu_n}{\partial n_n}\nablav \delta n_n\,,
\label{eq:euler_n}
\end{equation}
where the derivative has to be taken at the equilibrium density
$n_{n}$. In the liquid phase, neutrons and protons are coupled
by strong interactions, and in addition the protons feel an
acceleration due to the variation of the Coulomb potential,
$\delta V$. The corresponding two ($a = n,p$) Euler equations are
\begin{equation}
\partial_t \pv_a(\rv,t) = -\sum_{b=n,p}
\frac{\partial\mu_a}{\partial n_b}
\nablav \delta n_{b}(\rv,t) - \delta_{ap} \nablav \delta V(\rv,t) \,.
\label{eq:euler_np}
\end{equation}
Note that the relations between $\vv_a$ and $\pv_a$,
\Eqs{eq:entrainment_n} and (\ref{eq:entrainment_np}), are linearized,
too, i.e., it is sufficient to calculate $M$ at the equilibrium
densities.

The variation of the Coulomb potential depends itself on the variation
of the proton and electron densities. We assume that the electron
density follows instantaneously the motion of the protons, leading to
a screening of the Coulomb interaction. The corresponding modified
Poisson equation for the Coulomb potential reads
\begin{equation}
\Big(\nabla^2-\frac{1}{\lambda^2}\Big)\delta V(\rv,t) = 
  -4\pi e^2 \delta n_p\,,
\label{eq:poisson}
\end{equation}
where $\lambda$ is the Debye screening length given by
\begin{equation}
\frac{1}{\lambda^2} = 4\pi e^2 \frac{\partial n_e}{\partial \mu_e}\,.
\end{equation}

Let us now consider harmonic oscillations, i.e., all deviations from
equilibrium ($\phi_a$, $\pv_a$, $\vv_a$, $\delta n_a$, $\delta V$)
oscillate like $e^{-i\omega t}$. Therefore we can replace all time
derivatives $\partial_t$ by a factor $-i\omega$. Using
\Eqs{eq:velocitypotential} and (\ref{eq:continuity}), we can express
$\vv_a$ and $\delta n_a$ in terms of $\phi_a$ and obtain the following
equations for $\phi_a$ and $\delta V$:

\begin{romanlist}[(ii)]
\item In the gas phase without protons:
\begin{align}
\nabla^2 \phi_n &= -\frac{\omega^2}{u^2}\phi_n\,, \label{eq:hydrogas1}\\
\nabla^2 \delta V &= \frac{1}{\lambda^2} \delta  V\,, \label{eq:hydrogas2}
\end{align}
where
\begin{equation}
u^2 = \frac{n_n}{M_n}\,\frac{\partial \mu_n}{\partial n_n}
\end{equation}
denotes the square of the sound velocity. 
\item In the liquid phase with protons ($a = n,p$):
\begin{align}
\nabla^2\phi_a &= -\omega^2 \sum_{b=n,p} U^{-2}_{ab}\phi_b
  -2 i \omega U^{-2}_{ap}\delta V\,, \label{eq:hydroliquid1}\\
\nabla^2\delta V &= -\frac{i\omega}{2}\sum_{b=n,p} \frac{1}{\Lambda^2_b}
    \phi_b
  +\Big(\frac{1}{\lambda^2}+\frac{1}{\Lambda^2_p}\Big)\delta V\,, 
  \label{eq:hydroliquid2}
\end{align}
where we have defined the abbreviations
\begin{equation}
U^{-2}_{ab} = \sum_{c=n,p} M_{ac}\frac{1}{n_c}\,
  \frac{\partial n_c}{\partial \mu_b}\,,\qquad 
\frac{1}{\Lambda_a^2} = 4\pi e^2 \frac{\partial n_p}{\partial\mu_a}\,.
\end{equation}
Note that the eigenvalues of the matrix $U$ are the sound velocities
of the two eigenmodes in the uniform neutron-proton liquid without
Coulomb interaction.\cite{DiGallo2011}
\end{romanlist}
%%%%%%%%%%%%%%%%%%%%%%%%%%%%%%%%%%%%%%%%%%%%%%%%%%%%%%%%%%%%%%%%%%%%%%%%%%%%%%
\subsection{Boundary conditions}
\label{sec:boundary}
%%%%%%%%%%%%%%%%%%%%%%%%%%%%%%%%%%%%%%%%%%%%%%%%%%%%%%%%%%%%%%%%%%%%%%%%%%%%%%
As mentioned before, the equations of the preceding subsection are
valid inside each phase but not at the interface between the gas and
the liquid phase. At the interface, they have to be supplemented by
suitable boundary conditions. In \Ref{DiGallo2011}, we used very
simple boundary conditions: the pressure and the normal velocities had
to be equal on both sides of the interface. These conditions
correspond to an impermeable interface. In the present work, we will
improve the boundary conditions and allow for a neutron flux across
the interface, as in \Refs{Magierski2004,MagierskiBulgac2004}. Similar
boundary conditions were also given in \Ref{Lazarides2008} in a
completely different context, namely for an unpolarized Fermi gas
surrounded by a polarized one in an atom trap.

Let us consider the case of a surface parallel to the $xy$ plane,
separating the gas phase (1) from the liquid phase (2). Since there
are no protons in phase 1, the velocity of the surface is obviously
equal to the normal component of the proton velocity in phase 2,
$v_{zp2}$, and the displacement of the surface from its equilibrium
position is given by $\xi_z = v_{zp}/(-i\omega)$. If the normal
component of the neutron velocity is different from the velocity of
the surface, this means that neutrons cross the surface and pass from
one phase to the other. The requirement that the neutron current
leaving phase 1, $n_{n1}(v_{zn1}-v_{zp2})$, must be equal to the
neutron current entering phase 2, $n_{n2}(v_{zp2}-v_{zn2})$, gives our
first boundary condition:
\begin{equation}
\label{eq:boundarycross}
n_{n1}v_{zn1} = n_{n2} v_{zn2} - \Delta n_n v_{zp2}\,,
\end{equation}
where
\begin{equation}
\Delta n_n = n_{n2}-n_{n1}\,.
\end{equation}
%Note that, although the above arguments were stated for the case that
%phase 1 is situated below the surface and phase 2 above,
%\Eq{eq:boundarycross} is also valid in the opposite case. 
This equation can be rewritten in terms of the phases $\phi_a$ as
\begin{equation}
\label{eq:boundarycross1}
\frac{n_{n1}}{M_n}\partial_z \phi_{n1} = \sum_{a=n,p} (n_{n2}
M^{-1}_{na} - \Delta n_n M^{-1}_{pa})\partial_z\phi_{a2}\,.
\end{equation}
In our notation, the quantity $M_n$ belongs to the neutron gas and is
calculated at density $n_{n1}$, while $M_{ab}$ belongs to the
liquid phase and is calculated at densities $n_{n2}$ and
$n_{p2}$.

Furthermore, the fact that neutrons can cross the surface implies
that, as in equilibrium, the neutron chemical potentials on both sides
of the surface must be equal, i.e.,
\begin{equation}
\label{eq:boundarychemical}
\delta\mu_{n1} = \delta\mu_{n2}\,.
\end{equation}
If we express $\delta\mu_n$ in terms of derivatives
$\partial\mu_a/\partial n_b$ and the density variations $\delta n_b$
and use in each phase the equations of \Sec{sec:hydro},
\Eq{eq:boundarychemical} reduces to the very simple condition
\begin{equation}
\label{eq:boundarychemical1}
\phi_{n1} = \phi_{n2}\,.
\end{equation}

The condition $P_1 = P_2$ of equal pressures\footnote{To linear order
  in the velocities, the distinction between pressure and generalized
  pressure\cite{Prix2004,Prix2005} is irrelevant.} used in
\Ref{DiGallo2011} becomes more complicated once the surface tension is
included. The pressure difference is then given by the Young-Laplace
formula\cite{Landau6}
\begin{equation}
\label{eq:younglaplace}
P_2-P_1 = \sigma\Big(\frac{1}{R_1}+\frac{1}{R_2}\Big)\,,
\end{equation}
where $R_1$ and $R_2$ are the principal curvature radii of the
interface and the sign is such that the pressure is higher in the
phase having a convex surface. Since we consider here a surface that is flat 
in equilibrium, its curvature arises only from the displacement $\xi_z$
and can be expressed in terms of derivatives of $\phi_{a2}$. The
pressure difference on the left-hand side of \Eq{eq:younglaplace} is
related to the density oscillations, for instance we can write the
deviation of the pressure in the neutron gas from its equilibrium
value as $\delta P_1 = n_{n1}\delta\mu_{n1}$. Using again the
equations of \Sec{sec:hydro}, one can rewrite \Eq{eq:younglaplace} as
\begin{equation}
\label{eq:younglaplace1}
\omega\Big(\sum_{a=n,p}n_{a2}\phi_{a2} - n_{n1}\phi_{n1}\Big)
  +2i n_{p2} \delta V_2 = \pm
\frac{\sigma}{\omega}\sum_{a=n,p}M^{-1}_{pa}
  (\partial_x^2+\partial_y^2) \partial_z \phi_{a2}\,,
\end{equation}
where the upper (lower) sign is valid in the case that phase 2 is
situated above (below) phase 1.

Let us now turn to the boundary conditions related to the Coulomb
potential $V$. First of all, $V$ itself has to be continuous at the
surface, i.e.,
\begin{equation}
\label{eq:boundarycoulomb}
\delta V_1 = \delta V_2\,.
\end{equation}
This ensures that the component of the electric field tangential to
the surface is continuous, too.\cite{Jackson}

To linear order in the deviations, the charge of the protons in the
region between the unperturbed and the perturbed surface can be
considered as a surface charge density $e\,n_{p2}\xi_z$. This gives rise
to a discontinuity of the electric field normal to the
surface,\cite{Jackson}
\begin{equation}
\label{eq:boundaryeperp}
\delta E_{z1} - \delta E_{z2} = 4\pi en_{p2}\xi_z\,.
\end{equation}
Expressing again all quantities in terms of the potentials $\delta V$ and
$\phi_a$, we obtain our last boundary condition
\begin{equation}
\label{eq:boundaryeperp1}
-2i\omega\partial_z(\delta V_2-\delta V_1) =
  \sum_{a=n,p}\Omega_a^2\partial_z\phi_{a2}\,,
\end{equation}
with the abbreviation
\begin{equation}
\Omega_a^2 = 4\pi e^2n_{p2}M^{-1}_{pa}\,.
\end{equation}
%%%%%%%%%%%%%%%%%%%%%%%%%%%%%%%%%%%%%%%%%%%%%%%%%%%%%%%%%%%%%%%%%%%%%%%%%%%%%%
\subsection{Collective modes in a periodic slab structure (lasagne phase)}
\label{sec:modesinlasagne}
%%%%%%%%%%%%%%%%%%%%%%%%%%%%%%%%%%%%%%%%%%%%%%%%%%%%%%%%%%%%%%%%%%%%%%%%%%%%%%
We will now consider the simple case of a periodic structure of
slabs. We suppose that for $0<z<L_1$ we are in the gas phase (1), for
$L_1 < z < L=L_1+L_2$ we are in the liquid phase (2), and for $L<z<L+L_1$
we are again in the gas phase, and so on, as shown in
\Fig{fig:periodic}.

Since this system is translationally invariant in $x$ and $y$
directions, it is clear that the $x$ and $y$ dependence of the
eigenmodes (i.e., of the potentials $\phi_a$ and $\delta V$) is of the
form
\begin{equation}
\phi_a(\rv) = \phi_a(z) e^{i(q_xx+q_yy)}\,,\qquad \delta V(\rv) =
\delta V(z) e^{i(q_xx+q_yy)}\,.
\end{equation}
From the periodicity of the system in $z$ direction it follows that
the eigenmodes satisfy the Bloch conditions\cite{Ashcroft}
\begin{gather}
\phi_a(z+L) = \phi_a(z)e^{iq_z L}\,,\qquad
\delta V(z+L) = \delta V(z)e^{iq_z L}\,.
\label{eq:bloch}
\end{gather}
For given $\vek{q}$, these conditions can only be satisfied for some
discrete values of $\omega$. This is the excitation spectrum we are
looking for. Because of the Bloch conditions (\ref{eq:bloch}), it is
sufficient to solve the coupled differential equations of
\Sec{sec:hydro} in the regions (1) from 0 to $L_1$ and (2) from $L_1$
to $L$, together with the boundary conditions of \Sec{sec:boundary} at
$z = L_1$ and $L$.

%%%%%%%%%%%%%%%%%%%%%%%%%%%%%%%%%%%%%%%%%%%%%%%%%%%%%%%%%%%%%%%%%%%%%%%%%%%%%
\subsection{Numerical results}
\label{sec:results}
%%%%%%%%%%%%%%%%%%%%%%%%%%%%%%%%%%%%%%%%%%%%%%%%%%%%%%%%%%%%%%%%%%%%%%%%%%%%%%
Let us now investigate the resulting excitation spectrum for a
specific example. The values for the equilibrium quantities will be
taken from the work by Avancini et al.,\cite{Avancini2009} who have
studied the structure of pasta phases in a relativistic mean field
model. Our geometry corresponds to the lasagne phase, which has been
found in \Ref{Avancini2009} in the case of zero temperature and
$\beta$-equilibrium for baryon number densities $ 0.077$~fm$^{-3}
\lesssim n_B \lesssim 0.084$~fm$^{-3}$. For our example we have chosen
an intermediate density, $n_B = 0.08$~fm$^{-3}$, as in
\Ref{DiGallo2011}. The corresponding properties of the two phases (1)
and (2) are summarized in Table~\ref{tab:structure}.

The microscopic results for the equilibrium configuration will be used
to estimate the value of the surface tension, too.  While the surface
tension favors big structures, the Coulomb interaction favors small
ones.  In the case of the Lasagne geometry, the Coulomb and surface
energies per volume are given by
\begin{equation}
\epsilon_C+\epsilon_S = \frac{\pi}{6} e^2 n_e^2 L_1^2 + \frac{2\sigma}{L}\,,
\end{equation}
where $n_e$ is the electron number density, related to the proton
number density in phase 2, $n_{p2}$, by the requirement of charge
neutrality ($n_eL = n_{p2} L_2$), and $L = L_1+L_2$. The size $L$ of
the structure can be obtained by minimizing $\epsilon_C+\epsilon_S$
with respect to $L$, keeping the ratio $L_1/L_2$ fixed. Inversely, the
value of the surface tension can be estimated from the values of $L_1$
and $L_2$ given by the microscopic calculations,\cite{Avancini2009} as
\begin{equation}
\sigma = \frac{\pi}{6} e^2n_e^2\,L\,L_1^2\,.
\end{equation}
In principle the Coulomb interaction will result in slightly
non-uniform density distributions of the protons inside phase 2 and of
the electron gas. However, from the Coulomb potential calculated with
the uniform density distributions one can estimate that this effect is
weak and can be neglected in a first approximation.

%%%%%%%%%%%%%%%%%%%%%%%%%%%%%%%%%%%%%%%%%%%%%%%%%%%%%%%%%%%%%%%%%%%%%%%%
\begin{table}[pt]
\tbl{\label{tab:structure} Properties of the lasagne phase within
  the model by Avancini et al.\cite{Avancini2009} studied in our
  example. The average densities of the total system are given by $\bar{n}_n
  = (L_1/L) n_n^{(1)}+(L_2/L) n_n^{(2)}$ etc. Baryon density and proton
  fraction are defined as $n_B = n_n+n_p$ and $Y_p = n_p/n_n$,
  respectively.}
{\begin{tabular}{@{}c@{\hspace{1cm}}c@{\hspace{1cm}}c@{\hspace{1cm}}c@{\hspace{1cm}}c@{}} \toprule
                &            & slab (1)      & slab (2)      & total \\
\colrule
$L$             & (fm)       & 9.40\hph{00}  & 7.38\hph{00}  & 16.78\hph{00} \\
$n_n$           & (fm$^{-3}$) & 0.0701        & 0.0885       & \hph{0}0.0782 \\
$n_p$           & (fm$^{-3}$) & 0\hph{.0000}  & 0.0041       & \hph{0}0.0018 \\
$n_B = n_n+n_p$ &(fm$^{-3}$)  & 0.0701        & 0.0926       & \hph{0}0.0800 \\
$Y_p = n_p/n_B$ &            & 0\hph{.0000}  & 0.0447       & \hph{0}0.0227 \\
\botrule
\end{tabular}}
\end{table}
%%%%%%%%%%%%%%%%%%%%%%%%%%%%%%%%%%%%%%%%%%%%%%%%%%%%%%%%%%%%%%%%%%%%%%%%
The same remarks of caution as in \Ref{DiGallo2011} concerning the
dimensions of the structure, the coherence length and the validity of
the superfluid hydrodynamics approach (cf. discussion in \Sec{sec:ws})
are of course valid here and the interested reader is refered to that
paper.

\begin{figure}
\centerline{\includegraphics[width=12cm]{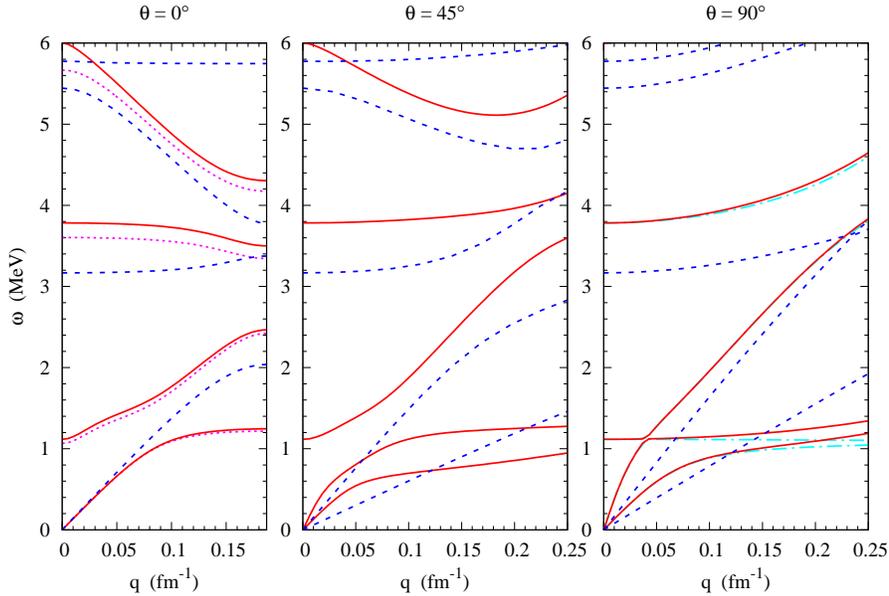}}
\caption{Dispersion relations of the modes propagating along the
  $z$-axis ($\theta = 0$, left), at an angle of $45^\circ$, and
  in the plane parallel to the slabs ($\theta = 90^\circ$, right). The
  plain red lines correspond to the present model and the dashed blue
  lines to the model of \Ref{DiGallo2011}, neglecting the Coulomb
  interaction and assuming an impermeable interface between the
  different phases. The dotted magenta lines indicate the results of
  the present model upon neglecting the microscopic entrainment effect
  and the dash-dotted turquoise line the result neglecting the surface
  tension.
  \label{fig:dispersion}}
\end{figure}

Let us now discuss the solutions for the energies $\omega$ and compare
the present model with that of \Ref{DiGallo2011}. In
Fig.~\ref{fig:dispersion}, the energies are shown
%%%%%%%%%%%%%%%%%%%%%%%%%%%%%%%%%%%%%%%%%%%%%%%%%%%%%%%%%%%%%%%%%%%%%%%%
%%%%%%%%%%%%%%%%%%%%%%%%%%%%%%%%%%%%%%%%%%%%%%%%%%%%%%%%%%%%%%%%%%%%%%%%
as functions of $q$ for three different angles $\theta$ between
$\vek{q}$ and the $z$ axis (i.e., $q_z = q \cos\theta$ and $q_\paral =
\sqrt{q_x^2+q_y^2} = q\sin\theta$). The left panel shows the
dispersion relation for waves propagating in $z$-direction,
i.e. perpendicular to the interfaces between the different slabs. We
observe an acoustic branch with an approximately linear dispersion law
\begin{equation}
\omega = u_1 q \label{acoustic3d}
\end{equation}
at low energies, and several optical branches with a finite energy for
$q = 0$, analogously to phonon branches in a crystal. The main differences
between the present model (solid red lines) and the previous
one\cite{DiGallo2011} (dashed blue lines) are as follows:
\begin{romanlist}[(ii)]
\item Here we use more realistic boundary conditions corresponding to
  a permeable interface between the two phases, whereas in
  \Ref{DiGallo2011} neutrons were not allowed to pass from one phase
  into the other.
\item We include the Coulomb interaction which was neglected in
  \Ref{DiGallo2011}.
\end{romanlist}
Actually, the change of the boundary conditions has only little
influence on the energy spectrum, whereas the Coulomb interaction
gives rise to an additional mode. Compared with \Ref{DiGallo2011}, we
have introduced the surface tension, too. The effect of the surface
tension is obviously vanishing for modes propagating perpendicular to
the interface and maximal for modes propagating along the layers. The
effect on the excitation energies is small, see the right panel of
\Fig{fig:dispersion}, where we display the result for $\sigma=0$ in
comparison with the complete calculation. On the other panels the
difference is too small to be seen and therefore not shown. The effect
of microscopic entrainment is weak, too, as can be seen from the left
panel.

Note that within the Wigner-Seitz approximation as discussed in
\Sec{sec:ws}, we would only obtain a discrete spectrum corresponding
to our spectrum in the case $q = 0$. The reason is that in this
approximation the coupling between cells is neglected, and thus each
cell has the same excitation spectrum. The degeneracy of the modes in
each cell is lifted by the coupling between cells, which gives rise to
a momentum dependent spectrum as obtained in our approach.

Let us now look at the cases $\theta = 45^\circ$ and $90^\circ$ shown
in the central and right panels of \Fig{fig:dispersion}. As in
\Ref{DiGallo2011}, we find now two acoustic modes. The energy of the
second acoustic mode can be approximately written as
\begin{equation}
\omega = u_2 q_\paral = u_2 q \sin\theta\,,
\label{eq:mode2d}
\end{equation}
which means that this mode propagates in a slab with almost no
coupling to neighboring slabs. 

\begin{figure}
\centerline{\includegraphics[width=8cm]{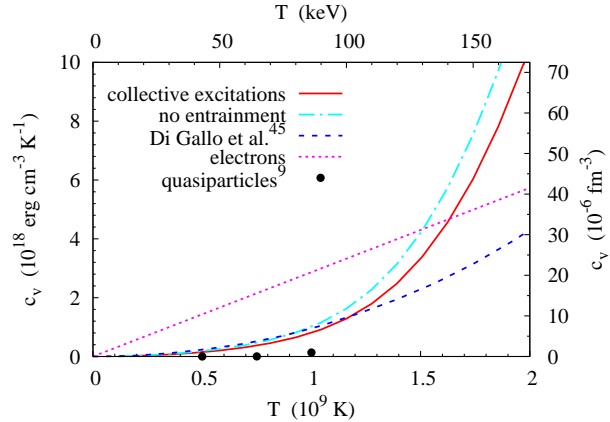}}
\caption{Specific heat as a function of temperature for $n_B = 0.08$~fm$^{-3}$.
  \label{fig:cv}}
\end{figure}

The two acoustic modes dominate largely the baryonic contributions to
the specific heat and exceed in particular the contributions from
superfluid neutron quasiparticles. 
The temperature dependence of the specific heat corresponding to the
modes shown in \Fig{fig:dispersion} is displayed in \Fig{fig:cv}.
%%%%%%%%%%%%%%%%%%%%%%%%%%%%%%%%%%%%%%%%%%%%%%%%%%%%%%%%%%%%%%%%%%%%%%%%
%%%%%%%%%%%%%%%%%%%%%%%%%%%%%%%%%%%%%%%%%%%%%%%%%%%%%%%%%%%%%%%%%%%%%%%%
We assume here that the temperature dependence of the mode spectrum
can be neglected, which should be true for
$T\ll\Delta_n,\Delta_p$. The electronic contribution $c_{v,e} =
\mu_e^2 T/3$ and the contribution from weakly paired neutron
quasiparticles\cite{FortinMargueron} are shown for comparison, too.

At very low temperatures, the specific heat of the collective excitations is
approximately given by\cite{DiGallo2011}
%%%%%%%%%%%%%
\begin{equation}
c_{v,\coll} \approx a T^2 + b T^3\,,\label{eq:cvapprox}
\end{equation}
%%%%%%%%%%%%%
with $a\approx 3\zeta(3)/\pi u_2^2L$ and $b\approx
2\pi^2/15u_1^3$. The $T^3$ term is the usual acoustic phonon, whereas
the ``two-dimensional'' mode discussed above gives rise to a
contribution increasing as $T^2$.\cite{DiGallo2011} We note, however,
that within our new model the deviations from this approximate formula
are stronger than in \Ref{DiGallo2011} for both acoustic modes: the
level repulsion due to avoided crossings let them deviate from the
purely linear behavior for much smaller momentum values. As a
consequence, the approximation (\ref{eq:cvapprox}) is valid only for
much smaller temperatures, where the electronic contribution to the
specific heat is the most important one. As can be seen from
\Fig{fig:cv}, the additional mode and the resulting flattening of the
acoustic modes increase the contribution to the specific heat.

\section{Summary and Conclusions}
In \Sec{sec:goldstone}, we reviewed a couple of existing approaches to
describe collective modes in the superfluid inner crust of neutron
stars. Considering only a uniform neutron gas within the QRPA, one
recovers the Goldstone mode, which is a longitudinal density wave (BA
sound) with linear dispersion relation at small $q$, as the lowest
lying collective excitation. For a realistic description of the
neutron star inner crust, however, its crystalline structure should be
considered, i.e. it should be described as a lattice of dense clusters
surrounded by a neutron gas. There are several approaches to treat
collective excitations of this non-uniform matter involving
different kinds of approximations.

In the standard hydrodynamic approach, only coarse-grained quantities
are considered, i.e., quantities averaged over many cells. This leads
to an effective low-energy theory, in which the superfluid collective
modes with linear dispersion relation are coupled to the dynamics of
the crystal lattice. However, this effective theory is only valid at
very large wavelengths (much larger than the periodicity of the
lattice, $L$), and the determination of its coefficients, especially
those involving the so-called entrainment, is difficult.

On the other hand, a QRPA calculation within the WS approximation
resolves the microscopic structure of each cell. However, it gives the
lowest lying collective mode at a finite energy. This is an artefact
of treating an isolated cell with a finite radius
$R_{\WS}$. Nevertheless the result is very important since it shows
that in a non-uniform system, superfluid hydrodynamics may be used at
a microscopic level (i.e., on length scales smaller than $R_{\WS}$) as
long as the coherence length is sufficiently short, i.e., pairing is
sufficiently strong.

In \Sec{sec:hydropasta}, we have presented the approach of superfluid
hydrodynamics at a microscopic level in order to develop a model for
the collective modes of large and intermediate wavelengths (i.e.,
larger than or comparable with $L$). In this model, hydrodynamic
equations in the clusters and in the neutron gas were matched at the
phase boundaries with boundary conditions allowing neutrons to pass
from one phase into the other. For simplicity, the model has so far
only been solved in the phase of plates. The most striking result is
the appearance of an approximately two-dimensional collective mode,
which gives a contribution to the specific heat proportional to $T^2$
(in contrast to the usual $T^3$ behavior) at low temperatures.

There remain many open questions. In particular, in order to better
understand the two-dimensional mode, we are developing an effective
theory corresponding to our model in the limit of small $q$. This
could help to clarify some questions concerning entrainment,
too,\cite{Magierski2004,MagierskiBulgac2004,Chamel2005,Chamel2012}
see the discussion in Sec.~\ref{sec:eft}. In addition, we plan to
apply our hydrodynamic model to other geometries such as the
crystalline phase or the phase of rods. 

The latter point is essential in order to make contact with
astrophysical observations, since it would allow one to determine heat
transport properties in the entire inner crust, necessary for modeling
neutron star thermal evolution. Properties of the crust influence the
cooling process mainly during the first 50-100 years, when the crust
stays hotter that the core which cools down very efficiently via
neutrino emission (see, e.g., \Ref{Yakovlev07}). Heat transport in the
crust is the key ingredient to explain the afterburst relaxation in
X-ray transients, too.\cite{Shternin2007,Brown09} As microscopic
ingredients for the models of the thermal relaxation of the crust, in
addition to the specific heat, thermal conductivity would be needed,
and to less extent neutrino emissivities. Although the contribution of
collective modes to heat transport properties exceeds largely that of
superfluid quasiparticles, it is to be expected that in almost the
whole crust electronic contributions remain dominant, except in
presence of a neutron star magnetic field of the order $10^{13}$ G or
higher,\cite{Aguilera08} which is the case for many observed neutron
stars. Therefore it would be interesting to extend the present model
to a magnetised environment, too.

\section*{Acknowledgements}
We thank Elias Khan for providing us with the numerical results of
\Ref{KhanSandulescu05}.
\appendix
%%%%%%%%%%%%%%%%%%%%%%%%%%%%%%%%%%%%%%%%%%%%%%%%%%%%%%%%%%%%%%%%%%%%%%%%%%%%%%
\section{Microscopic Input for the Model of Section \ref{sec:hydropasta}}
\label{sec:microscopicinput}
%%%%%%%%%%%%%%%%%%%%%%%%%%%%%%%%%%%%%%%%%%%%%%%%%%%%%%%%%%%%%%%%%%%%%%%%%%%%%%
As microscopic input, we need the equation of state, i.e., the
relation between the densities $n_a$ and the chemical potentials
$\mu_a$. As in \Ref{DiGallo2011}, in our concrete numerical examples,
we use the results of the work by Avancini et al.\cite{Avancini2009}
for the equilibrium configurations. They evaluate the structure of the
pasta phases for charge neutral matter in $\beta$ equilibrium using a
density dependent relativistic mean-field model, the DDH$\delta$ model
(originally called DDH$\rho\delta$), for the nuclear
interaction.\cite{Gaitanos04,Avancini04,Avancini2009} In order to be
consistent, we shall calculate the chemical potentials $\mu_a$ with
the same interaction. The entrainment coefficients within this model,
evaluated following the approach in \Refs{ComerJoynt2003,Gusakov2009},
are given by
\begin{equation}
M_{ab} = m^L_a\delta_{ab} + n_b\Big(\frac{\Sigma_R}{n_B} + 
  \frac{\Gamma^2_\omega}{m^2_\omega} + 
  \frac{t_at_b}{4}\, \frac{\Gamma^2_\rho}{m^2_\rho}\Big)\,,
\end{equation}
where $n_B = n_n+n_p$ denotes the baryon density, $t_n = -1$ and $t_p
= 1$. The Landau effective masses are denoted $m^L_a = \sqrt{p^2_{Fa}
  + m_a^{D\,2}}$, not to be confused with the Dirac effective masses
$m^D_a$. The Fermi momenta are represented by $p_{Fa}$ and $\Gamma_i =
g_i h_i(n_B)$ are the meson-nucleon coupling constants of the
DDH$\delta$ model, see \Ref{Avancini2009}. The term $\Sigma_R$, not
present in the expressions given in \Refs{ComerJoynt2003,Gusakov2009},
arises from the density dependence of the coupling constants and
appears in the Landau effective masses, too,
\begin{equation}
m^L_a = \mu_a - \frac{\Gamma^2_\omega}{m^2_\omega} n_B - t_a
\frac{\Gamma^2_\rho}{4 m_\rho^2} (n_p- n_n) - \Sigma_R\,.  
\end{equation}
and an explicit expression can be found in \Ref{Avancini2009}. 
%%%%%%%%%%%%%%%%%%%%%%%%%%%%%%%%%%%%%%%%%%%%%%%%%%%%%%%%%%%%%%%%%%%%%%%%%%%%%%

\end{document}